# A Monster in the early Universe

Chris Willott

Canadian Astronomy Data Centre
Herzberg Institute of Astrophysics
National Research Council Canada
Victoria, British Columbia V9E 2E7, Canada
e-mail: chris.willott@nrc-cnrc.gc.ca

**The most distant quasar yet discovered sets constraints on the formation mechanism of black holes. Its light spectrum has tantalizing features that are expected to be observed before the reionization epoch ended.**

The finite speed of light elevates astronomers' telescopes to powerful time machines that allow us to look back into the early history of the cosmos. Light travels almost unimpeded through the Universe for billions of years before reaching Earth, its wavelength stretched by the expansion of space itself. The quest for the most distant — and hence earliest — objects in the Universe, to study the initial phase of galaxy formation, is at the forefront of observational cosmology. On page 616 of this issue, Mortlock *et al.*[1] present the discovery of the most distant accreting supermassive black hole, or quasar, found so far — at a redshift of 7.085, or just 770 million years after the Big Bang. The quasar is a monster: a black hole with a mass two billion times that of the Sun, accreting gas at the maximum rate allowed by the laws of physics. The discovery is significant both for the existence of these supermassive black holes at early times and for the information that the quasar light provides on the state of the surrounding Universe.

All massive galaxies, including our own, contain central black holes with mass greater than one million times that of the Sun[2]. Black holes grow by accreting matter and by merging with other black holes. The total cosmological accretion and merging can be observationally constrained by the luminosity output of quasars and by hierarchical structure formation, respectively. But how these black holes got started is still unknown. We know that massive stars leave relic black holes after they explode as supernovae and that the primordial element abundance in the early Universe produced a higher fraction of massive stars than form today. Copious accretion and merging would be required to turn these 10–100-solar-mass black holes into the monsters observed at the centres of galaxies. Another possibility is that the star-formation step was skipped altogether as a million-solar-mass cloud of primordial gas collapsed without fragmenting, forming a black hole directly[3].

How to choose between these scenarios? That's where the new quasar discovery[1] comes in. Black-hole growth by accretion is limited by radiation pressure, and so is an exponential process much like bacterial population growth by cell division. The mass-doubling timescale is about 50 million years[4]. At 770 million years after the Big Bang, Mortlock and colleagues' quasar can have undergone a maximum of just 15 doublings since the beginning of time. Given that the quasar has a mass two billion times that of the Sun, this simple argument suggests either that the quasar had a single progenitor with a mass at least half a million times that of the Sun, or that it resulted from the merging of several thousand massive-star remnants at the centre of its host galaxy. These constraints can be loosened by invoking mergers with black holes in other galactic nuclei or periods of extreme accretion[5], but it is safe to say that the existence of this quasar will be giving some theorists sleepless nights.

The new quasar is observed at a particularly interesting era of the Universe. Cosmic reionization, the transition from a mostly neutral to mostly ionized intergalactic medium (IGM), occurred sometime between redshifts 6 and 15. Reionization depends on properties of the first generation of star and galaxy formation, and is an important probe of the early Universe. Quasars are used as background light sources to study the absorption of Lyman-α photons by neutral hydrogen between the quasars and Earth. A range of tests have been devised to determine the neutral-hydrogen fraction in the IGM, both close to quasars and in the vast, mostly empty, intervening space. These analyses have shown that, at redshifts below 6, the IGM hydrogen is mostly ionized, although a significant neutral fraction exists[6] in some directions at redshifts 6 to 6.4. This, combined with observations[7] of relic radiation from the Big Bang (the cosmic microwave background), suggests that most reionization occurred at redshifts higher than 6.4 — less than 880 million years after the Big Bang.

So what does the new quasar tell us about the ionization state of the IGM at redshifts 6 to 7.085? The quasar's light spectrum, obtained using the Very Large Telescope and the Gemini North Telescope, shows features known as Gunn–Peterson troughs spanning a wide wavelength range, placing a lower limit on the neutral-hydrogen fraction in the IGM from redshift 6 to 7 of about $10^{-3}$. There are two observational signatures of the IGM ionization state close to the quasar that are sensitive to large neutral fractions — the size of the quasar's ionized near-zone[8] and the damping wing of Lyman-α absorption from the IGM beyond this zone[9] (Fig. 1). For both these effects, the spectrum of the new quasar shows tantalizing differences from redshift-6 quasars, suggesting a high neutral-hydrogen fraction (greater than 0.1) in the IGM close to the quasar and that the quasar is in the midst of the reionization era.

The quasar was first identified in images obtained with the United Kingdom Infrared Telescope (UKIRT) Infrared Deep Sky Survey, and is a triumph for this project, which was initiated in 2005. It illustrates the power of dedicated wide-field surveys on moderate-sized telescopes, and marks the transition from the optical to near-infrared wavelength regimes in the search for the highest-redshift quasars.

There will be many follow-up observations of this quasar to investigate reionization constraints and study the black hole and galaxy it resides in. But one object does not always tell the whole story, so it will be important to find more quasars at redshifts above 7, and to push discoveries towards the early phase of reionization at redshift 10. This will be challenging, as illustrated by the marked decrease in the abundance of galaxies at this redshift[10], but is possible with a wide-field, near-infrared space telescope.

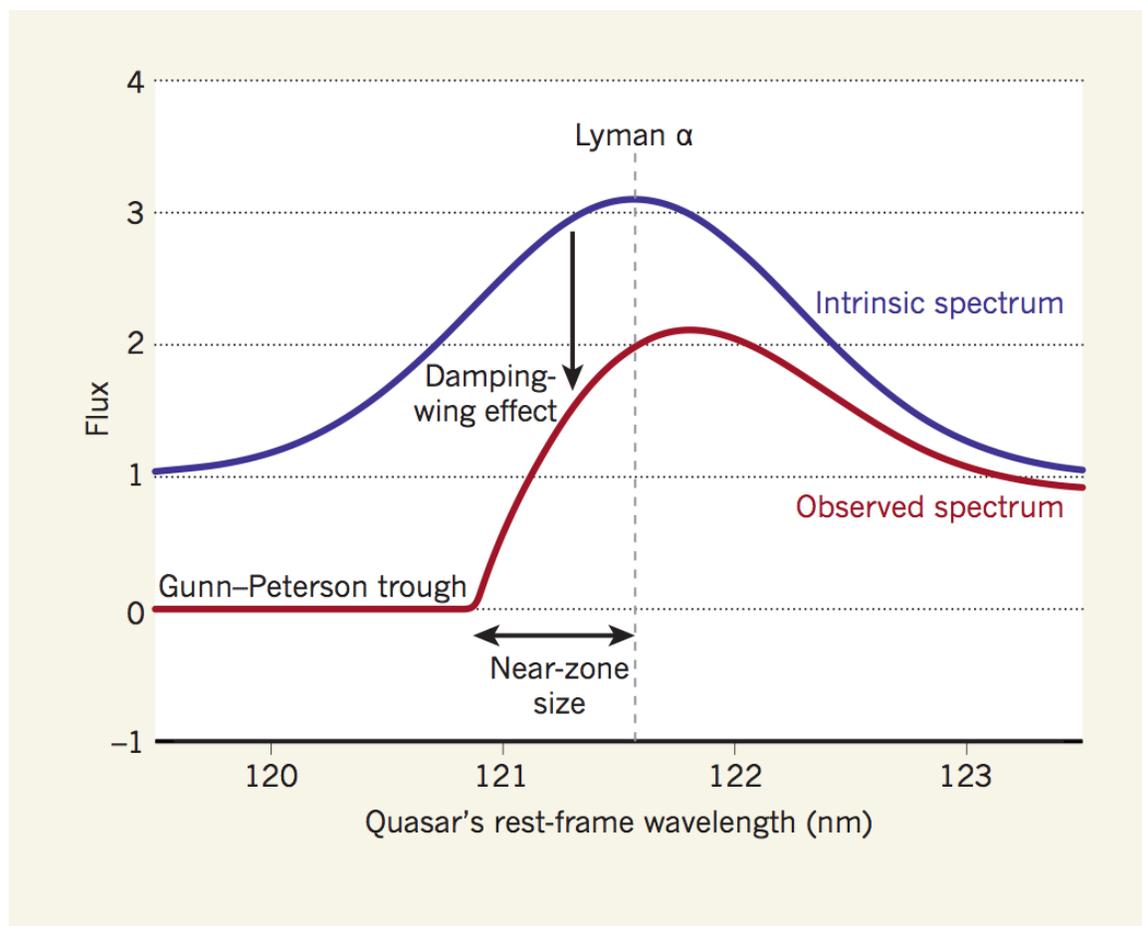

**Figure 1. Lyman-α emission line spectrum**. The blue curve is the intrinsic quasar emission of a hypothetical redshift-7 quasar embedded in a neutral intergalactic medium (IGM). The red curve is that observed after absorption by neutral hydrogen in the IGM. The Gunn–Peterson trough is due to IGM absorption along the line of sight to the quasar. The damping-wing effect[9] is due to absorption from this same region, but, unlike the Gunn–Peterson trough, requires a very high neutral-hydrogen fraction. The transmitted flux in the quasar near-zone is due to a self-ionized region

around the quasar. The size of this region depends on several factors, one being the neutral fraction of the IGM before the quasar turned on[8]. The quasar discovered by Mortlock et al.[1] shows a small near-zone and a likely damping wing, implying a high neutral-hydrogen fraction at redshift 7.08.